\documentclass{aa}
\usepackage{graphicx}
\usepackage{txfonts}
\usepackage{natbib}
\bibpunct{(}{)}{;}{a}{}{,} 

\begin{document}
\title{IGR~J18483$-$0311: a new intermediate supergiant fast X-ray transient\thanks{Based on
    observations carried out at the European Southern Observatory
    under the Target of Opportunity programme ID~078.D-0268.}}

\author{F. Rahoui
  \inst{1,2}
  \and
  S. Chaty
  \inst{1}
}

\offprints{F. Rahoui}

\institute{Laboratoire AIM, CEA/DSM - CNRS - Universit\'e Paris Diderot,
  Irfu/Service d'Astrophysique, B\^at. 709, CEA/Saclay, F-91191
  Gif-sur-Yvette, France, \email{frahoui, chaty@cea.fr}
  \and
  European Southern Observatory, Alonso de C\'ordova 3107,
  Vitacura, Santiago de Chile\\
}

\date{Received; accepted}


\abstract
    {IGR~J18483$-$0311 is a high-mass X-ray binary
      recently discovered by \textit{INTEGRAL}. Its periodic
      fast X-ray transient activity and its position in the Corbet
      diagram - although ambiguous - led to the conclusion 
      that the source was a likely Be/X-ray binary
      (BeXB), even if a
      supergiant fast X-ray transient (SFXT) nature could not be excluded.} 
    {We aimed at identifying the companion star of IGR~J18483$-$0311 to
      discriminate between the BeXB and the SFXT nature of the source.} 
    {Optical and near-infrared photometry, as well as
      near-infrared spectroscopy of the companion star were performed to identify its spectral type. We
      also assembled and fitted its broad-band spectral energy distribution to
      derive its physical parameters.}
    {We show that the companion star of IGR~J18483$-$0311 is an early-B
      supergiant, likely a B0.5Ia, and that its distance is about 3-4~kpc.}
    {The early-B supergiant nature of its companion star, as well as
      its fast X-ray transient activity point towards an SFXT nature
      of IGR~J18483$-$0311. Nevertheless, the long duration and the periodicity of its
      outbursts, as well as its high level of quiescence, are 
      consistent with IGR~J18483$-$0311 being an intermediate SFXT,
      in between classical supergiant X-ray binaries (SGXBs) 
      characterised by small and circular orbits, and classical SFXTs with
      large and eccentric orbits.}
    

\maketitle
%

\section{Introduction}

High-mass X-ray binaries (HMXBs) are X-ray sources for which 
high-energy emission stems from accretion onto a compact object
of material coming from a massive
companion star. Until recently, the majority of known HMXBs
were Be/X-ray binaries (BeXBs), and just a few were
supergiant X-ray binaries (SGXBs).
The launch of the \textit{INTErnational Gamma-Ray Astrophysics Laboratory}
\citep[\textit{INTEGRAL}, ][]{2003Winkler} in October 2002 completely
changed the situation, as many more HMXBs whose companion stars
are supergiants were discovered during the monitoring
of the Galactic centre and the Galactic plane using the onboard
IBIS/ISGRI instruments \citep{2003Ubertini, 2003Lebrun}. Most of
these sources are reported in
\citet{2007Bird} and \citet{2007Bodaghee}, and their studies
reveal two main features which were not present on previously
known SGXBs: first, many of them exhibit a
considerable intrinsic absorption, far above the interstellar level;  
second, some of these new sources revealed a transitory nature and
occasionally exhibit a fast X-ray transient activity lasting a few hours.
It then appears that the \textit{INTEGRAL} supergiant HMXBs can be
classified in two classes: one
class of considerably obscured persistent sources \citep[see e.g. ][]{2006Chaty} and another of
transitory sources called supergiant fast X-ray transients \citep[SFXTs,][]{2006Negueruelaa}.
\newline

IGR~J18483$-$0311 was discovered during
observations performed with \textit{INTEGRAL} in 2003 April 23-28
\citep{2003Chernyakova}, and it was found to have average fluxes
of about 10~mCrab and 5~mCrab in the 15$-$40~keV and 40$-$100~keV bands,
respectively. \citet{2004Molkov} also detected the source in 2003 March-May during a
survey of the Sagittarius arm tangent with \textit{INTEGRAL}, and
gave average fluxes of about 4.3~mCrab and 3.9~mCrab in the 18$-$60~keV
and the 60$-$120~keV bands, respectively. Moreover, analysing archival
data from the \textit{Rossi X-ray Timing
  Explorer} (\textit{RXTE}) All-Sky Monitor (ASM),
\citet{2006Levine} reported a 18.55~days periodicity of the
lightcurve of IGR~J18483$-$0311, likely corresponding to its orbital
period.

With archival data from observations performed with \textit{INTEGRAL}
from 2003 May to 2006 April, \citet{2007Sguera} found five
new outbursts from IGR~J18483$-$0311 whose activity lasted no more
than a few days, and characterised by fast flares lasting a few
hours. They fitted its 3$-$50~keV
spectrum with an absorbed cut-off power law, and derived
$N_{\rm H}\sim9\times10^{22}$~cm$^{-{2}}$ (well above the
galactic column density in the line of sight of about
$1.6\times10^{22}$~cm$^{-{2}}$), $\Gamma\sim1.4$, and
$\textrm{E}_\textrm{c}\sim22$~keV. They also showed that all
the outbursts are well-fitted by an absorbed bremsstrahlung whose
parameters are $N_{\rm H}\sim7.5\times10^{22}$~cm$^{-{2}}$
and $\textrm{kT}\sim21.5$~keV. They reported a periodicity
in the long term ISGRI light curve of about 18.52~days - confirming
the result of
\citet{2006Levine} - and argued it is most likely an orbital
period. They also derived a periodicity of about 21.05~s in the
JEM-X light curve of the first outburst, which is likely a
neutron star pulse period. Finally, from archival \textit{Swift}
observations, the authors obtained a very accurate position of
IGR~J18483$-$0311, which allowed them to pinpoint its
USNO~B1.0~0868$-$0478906 and 2MASS~J18481720$-$0310168 counterparts. They found 
that the magnitudes were typical
of an absorbed massive late O/early B star which, along with the
position of the source in the Corbet diagram as well as the
periodicity of its high-energy activity, led them to conclude
that IGR~J18483$-$0311 was likely a BeXB, without
excluding the possibility of an SFXT nature.

\citet{2008Chaty} assembled the broad-band spectral energy
distribution (SED) of
the companion star of IGR~J18483$-$0311 from 0.7~$\mu$m to 8~$\mu$m, 
with near-infrared (NIR) photometric observations 
performed at the ESO/NTT using the SofI instrument, as well as
archival optical data from the USNO-A.2 catalogue, and mid-infrared
(MIR) data from the GLIMPSE survey. They fitted it with a combination of two absorbed
black bodies and showed that the best fit was consistent with the
companion star being a heavily absorbed B star, confirming the
HMXB nature of IGR~J18483$-$0311. At last, \citet{2008Masetti} performed optical spectroscopy of the
companion star of IGR~J18483$-$0311, and concluded that it was an O/B
giant star because of the large equivalent width of the H$_{\rm \alpha}$
emission line.
\newline

In this paper, we report optical and NIR observations of
IGR~J18483$-$0311 performed in service mode at the ESO/NTT through our Target of
Opportunity program ID~078.D-0268 (P.I. S. Chaty). These
observations aimed at constraining the nature of its companion
star and of the binary system. In Sect. 2, we describe the optical/NIR photometric and
spectroscopic observations, as well as their reduction. In Sect. 3, we
present the results and show the broad-band SED of the companion star of IGR~J18483$-$0311. 
We finally discuss the outcomes in Sect. 4.

\section{Observations and data analysis}

\subsection{Optical and NIR photometry}

In 2007 March 24, we performed optical photometry of IGR~J18483$-$0311
in \textit{U}, \textit{B}, \textit{V}, \textit{R}, and
\textit{I} bands with the imager SUSI2 installed at ESO/NTT. We used
the large field of view (5\farcm5$\times$5\farcm5), and the plate scale was 0\farcs16 
per pixel. The exposure time was 60~s in all filters for each
exposure. We also observed four photometric standard stars from the
optical standard catalogue of \citet{1992Landolt}: PG1657+056,
PG1657+056A, PG1657+056B, and PG1657+056C.

NIR photometry was carried out in \textit{J}, \textit{H}, and \textit{Ks} bands
with the spectro-imager SofI at ESO/NTT. We used the
large field of view (4\farcm92$\times$4\farcm92), and the plate
scale was 0\farcs288 per pixel. In each filter, the observations were
obtained by repeating a set of images with 9 different 30$\arcsec$
offset positions to subtract the sky emission, and the exposure
time was 1.2~s for each position. We also observed three
photometric standard stars from the faint NIR standard star
catalogue of \citet{1998Persson}: sj9116, sj9150, and sj9155.

The photometric data were reduced using \textit{IRAF} by performing
flatfielding and bad pixels
correction, as well as bias subtraction in the optical bands, and
crosstalk correction and sky subtraction in the NIR. After the reduction,
all the images were aligned and combined. We then
carried out aperture photometry on the optical images. In the NIR,
we used the software \textit{StarFinder}, part of the
\textit{Scisoft} package from ESO, well-suited for point-source
photometry in crowded fields. The instrumental magnitudes were then transformed to apparent
magnitudes using the zero-point, the extinction coefficient, and the color
term corrections computed by linear fits using the standard stars.
The optical and NIR magnitudes are listed in Table~1. It is worth noting
that our optical magnitudes are about 1~magnitude lower than those given in the USNO
catalogue, this difference probably due to the strong
uncertainty ($\geq1$~magnitude) of the USNO magnitudes for very faint
sources ($\ge$17 in the \textit{B} band).

\subsection{NIR spectroscopy}

In 2007 March 18-19, we performed NIR low-resolution (R$\sim$1000)
spectroscopy with SofI between 0.9~$\mu$m and 2.5~$\mu$m, using the blue and red
grisms and the 1$\arcsec$ slit. Twelve spectra were taken in each
grism, half of them on the
source and the other half with an offset of 30$\arcsec$ and allowing for sky
emission subtraction. The exposure time of each spectrum was set to
60~s. We also observed in the same conditions HIP96265, a G3V telluric
standard star, for telluric features correction.

The NIR spectra were reduced with \textit{IRAF} by performing crosstalk correction,
flatfielding, sky subtraction, and bad pixels correction.
The target spectra were then extracted, wavelength calibrated by
extraction of a xenon arc with the same setup, and finally combined. 
The telluric features were corrected by dividing the final target
spectrum by the telluric standard. To avoid line contamination by
the standard star, we finally multiplied the corrected 
spectrum by a solar spectrum provided by the National Solar Observatory
(NSO), and modified using an \textit{IRAF} routine
developed by \citet{1996Maiolino}. The routine corrects the solar
spectrum for airmass, heliocentric, and
rotational velocity to match the standard star properties, and
smoothes the spectrum to rebin it to the instrument resolution. 
The final spectrum of the companion star of IGR~J18483$-$0311 - obtained
by combining the blue and red grisms - is displayed in Fig~1.

\begin{table*}
  \caption{Optical and NIR magnitudes (Mag) of 
    IGR~J18483$-$0311 obtained with SUSI2 and SofI. We also give the
    zero point $Z_{\rm p}$ in each filter.} 
  $$
  \begin{array}{c|c c c c c c c c}
    \hline
    \hline
    \textrm{Filters}&U&B&V&R&I&J&H&Ks\\
    \hline
    Z_{\rm p}&23.279\pm0.023&24.837\pm0.030&24.423\pm0.018&24.480\pm0.013&22.707\pm0.061&23.059\pm0.020&22.814\pm0.013&22.211\pm0.019\\
    \hline
    \textrm{Mag}&>23.702&>25.162&21.884\pm0.313&17.888\pm0.041&14.382\pm0.061&10.840\pm0.028&9.376\pm0.018&8.472\pm0.023\\
    \hline
  \end{array} 
  $$
\end{table*}  

\begin{figure*}
  \includegraphics[width=17.5cm, height=8cm]{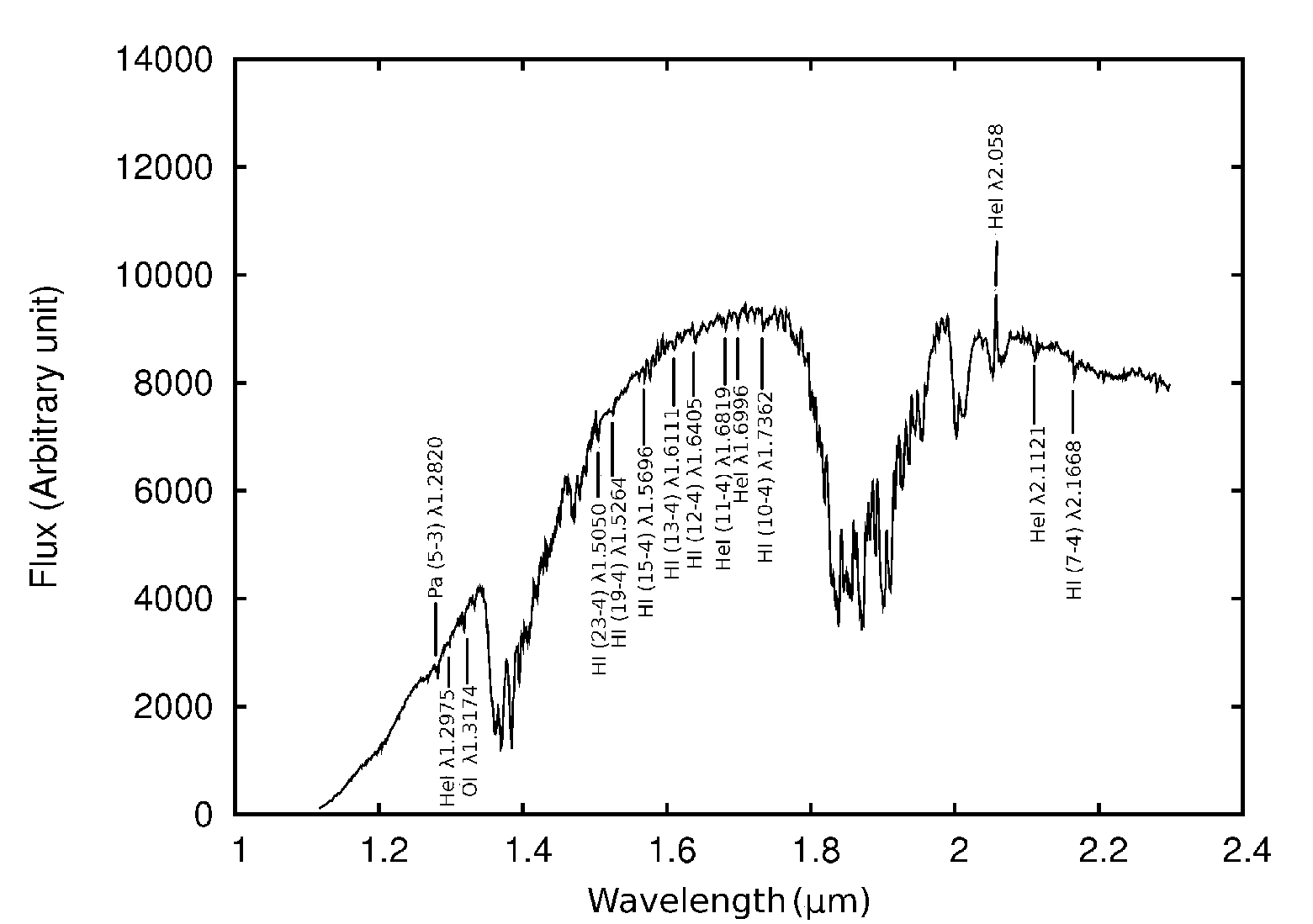}
  \caption{1.12~$\mu$m to 2.3~$\mu$m low-resolution spectrum of the companion star of
    IGR~18483$-$0311, obtained with the blue and red grisms of
    SofI.}
\end{figure*}

\section{Results}

\subsection{Spectral classification}

To identify the spectral type of the companion star of
IGR~J18483$-$0311, we compared our spectrum with spectra from the
\textit{H} and \textit{K} bands hot stars classifications given in
\citet{1996Hanson,2005Hanson} and the \textit{K} band
spectroscopic classification given in \citet{2000Clark}. In the
spectrum displayed in Fig~1, we marked all the detected
features (listed in Table 2), most of them due to
\ion{H}{i} and \ion{He}{i}. In the \textit{H} band spectrum, we also detected
the \ion{Pa}{$\beta$}~(5$-$3) feature at 1.2822~$\mu$m and the \ion{O}{i}
feature at 1.3170~$\mu$m. 

The presence of the \ion{Br}{$\gamma$}~(7$-$4) line in absorption at 2.1668~$\mu$m
and of several \ion{He}{i} lines is typical of a normal 09-B3
massive star \citep{2000Clark}. This therefore excludes the
system to be a BeXB. Moreover, the presence of the \ion{He}{i} line at
2.0580~$\mu$m in emission and the \ion{He}{i} line at 2.1121~$\mu$m in
absorption is typical of early B0-B2 supergiants. 
Comparing the relative strengths of the \ion{He}{i}
lines at 1.6996~$\mu$m, 2.0580 and 2.1121~$\mu$m, and of the \ion{H}{i} lines at
1.6819~$\mu$m, 1.7362~$\mu$m, and 2.1668~$\mu$m, with the relative strengths of the same lines in
\citet{1996Hanson,2005Hanson}, we find that the star is likely a
B0.5Ia supergiant.

\begin{table}
  \caption{Spectroscopic features we detected in the blue and red
    spectra of the companion star of IGR~J18483$-$0311. We also give
    their laboratory and fitted wavelengths in $\mu$m ($\lambda$ and
    $\lambda_{\rm fit}$, respectively), their
    equivalent width ($\mathring{W}$) and their full-width at
    half-maximum (\textit{FWHM}), in \AA.}
  $$
  \begin{array}{c c c c c}
    \hline
    \hline
    \textrm{Identification}&\lambda&\lambda_{\rm fit}&\mathring{W}&FWHM\\
    \hline
    \textrm{Pa}~(5$-$3)&1.2822&1.2820&1.013&8.351\\
    \hline
    \ion{He}{i}&1.2976&1.2975&0.765&20.97\\
    \hline
    \ion{O}{i}&1.3165&1.3170&1.700&17.4\\
    \hline
    \textrm{Br}~(23$-$4)&1.5043&1.5050&0.949&22.53\\
    \hline
    \textrm{Br}~(19$-$4)&1.5265&1.5264&1.018&41.25\\
    \hline
    \textrm{Br}~(15$-$4)&1.5705&1.5698&0.856&23.40\\
    \hline
    \textrm{Br}~(13$-$4)&1.6114&1.6111&1.049&37.56\\
    \hline
    \textrm{Br}~(12$-$4)&1.6412&1.6405&0.641&25.94\\
    \hline
    \textrm{Br}~(11$-$4)&1.6811&1.6819&0.723&27.25\\
    \hline
    \ion{He}{i}&1.7007&1.6996&1.053&32.72\\
    \hline
    \textrm{Br}~(10$-$4)&1.7367&1.7362&0.813&33.26\\
    \hline
    \ion{He}{i}&2.0587&2.0580&-5.794&32.16\\
    \hline
    \ion{He}{i}&2.1126&2.1121&1.168&30.67\\
    \hline
    \textrm{Br}~(7$-$4)&2.1661&2.1668&1.151&23.17\\
    \hline
  \end{array} 
  $$
\end{table}  

\subsection{Spectral energy distribution}

To assemble the broad-band SED of 
IGR~J18483$-$0311, we added its archival MIR fluxes from the
\textit{Spitzer}'s Galactic Legacy Infrared Mid-Plane Survey
Extraordinaire \citep[GLIMPSE,][]{2003Benjamin} of the Galactic plane
($|b|\leq1^\circ$ and between \textit{l}=10$^\circ$ and
\textit{l}=65$^\circ$, on both sides of the Galactic centre)
performed with the \textit{Spitzer Space Telescope} with the IRAC
camera in four bands, 3.60$\pm$0.38~$\mu$m, 4.50$\pm$0.51~$\mu$m,
5.80$\pm$0.73~$\mu$m, and 8.00$\pm$1.43~$\mu$m. The GLIMPSE fluxes of the
companion star of IGR~J18483-0311 are listed in \citet{2008Chaty}.
We then fitted the SED (using a $\chi^2$ minimisation) with an
absorbed black body representing the companion
star emission, following \citet{2008Rahoui}. 
We added to the flux uncertainties a 5\% systematic error in each NIR
band, and a 2\% systematic error in each IRAC band, as given in the
SofI\footnote{http://www.ls.eso.org/docs/LSO-MAN-ESO-40100-0004/LSO-MAN-ESO-40100-0004.pdf} 
and the IRAC\footnote{http://ssc.spitzer.caltech.edu/documents/som/som8.0.irac.pdf}
manuals. The best-fitted SED and parameters are displayed in Fig~2 and
listed in Table~3, respectively.
\begin{figure}
  \includegraphics[width=8.2cm]{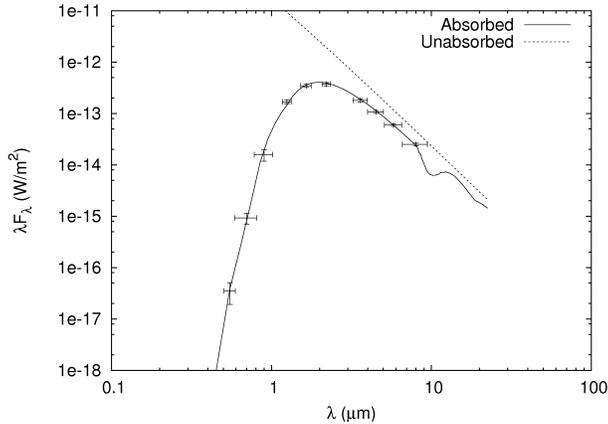}
  \caption{Broad-band SED of 
    IGR~J18483$-$0311. Assembled with optical, near-, and
    mid-infrared data from SUSI2, SofI, and IRAC, respectively.}
\end{figure}
\newbox\hautbox \setbox\hautbox=\hbox{\vphantom{\rule[-0.15cm]{0cm}{0.45cm}}}
\begin{table}
  \caption{Best-fitted parameters of the SED of IGR~J18483$-$0311. We give the
    total galactic absorption in the line of sight due to \ion{H}{i}
    ($A_{\rm \ion{H}{i}}$) and H$_2$ ($A_{\rm H_2}$),
    the absorption of the source in the X-ray domain
    $A_{\rm x}$, and then the parameters: the
    optical intrinsic absorption 
    $A_{\rm v}$, the temperature
    $T_{\rm \ast}$ and the $R_{\rm \ast}/D_{\rm \ast}$
    radius-to-distance ratio of the
    star. The absorptions are given in magnitudes.}
  $$ 
  \begin{array}{@{\usebox{\hautbox}}cccccccc}
    \hline
    \hline
    A_{\rm \ion{H}{i}}&A_{\rm H_2}&A_{\rm x}&A_{\rm v}&T_{\rm
      \ast}[\Delta{T}_{\rm \ast}](\textrm{K})&R_{\rm \ast}/D_{\rm \ast}\\
    \hline
    7.9&9.6&48.1&15.7_{-0.3}^{+0.2}&24600[17500\,-\,36800]&1.94_{-0.41}^{+0.39}\times10^{-{10}}\\
    \hline
  \end{array}
  $$ 
\end{table}      

\subsection{Distance}

The best-fitted temperature of the companion star of IGR~18483$-$0311
($T_{\rm \ast}=24600$~K) is consistent with the
star being a B0.5Ia supergiant, following the temperature
calibration given in \citet{2008Searle}. In their paper, they
also give the expected unabsorbed absolute magnitude of a B0.5Ia
supergiant star, $M_{\rm v}=-6.5$, as well as its
expected radius, $R_{\rm e}=33.8{\textrm{R}_{\rm \odot}}$. Using these
two parameters, as well as $A_{\rm v}$
and $R_{\rm \ast}/D_{\rm \ast}$ that we derived
from our fit, we can assess the distance of the
source with one of the following relations: 
$D_\ast\,=\,10^{\textrm{\normalsize $0.2(m_{\rm v}-A_{\rm v}-M_{\rm v}+5)$}}$ (in pc), or 
$D_\ast\,=\,$ \normalsize $\frac{R_{\rm e}}{R_{\rm \ast}/D_{\rm \ast}}$ (in $\textrm{in R}_{\rm \odot}$), 
where $m_{\rm v}$ is the optical apparent magnitude of the source. Both relations give $D_{\rm \ast}=3.44$~kpc and
$D_{\rm \ast}=3.93$~kpc, respectively. Both values are
consistent within the uncertainties and we therefore conclude
that the distance of IGR~J18483$-$0311 is about 3-4~kpc.

\section{Discussion and conclusions}

The identification of its companion star as a B0.5Ia supergiant,
as well as its fast X-ray transient activity, are consistent with IGR~J18483$-$0311 being an
SFXT. Nevertheless, as already pointed out in \citet{2007Sguera},
the IGR~J18483$-$0311 X-ray behaviour appears unusual. Indeed, whereas outbursts last a few
hours and $L_{\rm max}$/$L_{\rm min}\sim10^4$ for typical SFXTs, the outbursts
of IGR~J18483$-$0311 last a few days and its $L_{\rm max}$/$L_{\rm min}$ ratio is
$\sim\,10^3$ (meaning that its quiescence is higher). Moreover,
\citet{2006Levine} and \citet{2007Sguera} reported a 18.52~days
orbital period, which is quite low compared to what is expected for
classical SFXTs, with large and eccentric orbits. 

Recently, several authors pointed out the importance of clumpy winds
to explain the SFXT behaviour \citep{2005Zand, 2007Leyder, 2007Walter,
  2008Negueruela}. They argue that the flares are created by the
interaction of the compact object with dense clumps (created in the
stellar wind of the companion star), the frequency of the
flares depending mainly on the geometry of the system. On
the contrary, the quiescent emission would be due to the accretion
onto the compact object of diluted inter-clumps medium, which
would explain the very low quiescence exhibited
 by classical SFXTs. Moreover, by classifying twelve SFXTs
in function of the duration and the frequency of their outbursts,
as well as their $L_{\rm max}$/$L_{\rm min}$ ratio, 
\citet{2007Walter} showed the existence of intermediate SFXTs,
characterised by a lower ratio and longer flares. Finally,
\citet{2008Negueruela} proposed a general scheme to explain the
emission of both SGXBs and SFXTs. The authors argued about
the existence of a zone around the companion supergiant star (radius $\le\,2R_{\rm \ast}$), in which
the clump density is very high, and another one (radius $\ge\,3R_{\rm \ast}$) in which it is low. In
function of the orbital parameters of the compact object, the system
could be a classical SGXB (small and
circular orbit inside the zone of high clump density), a classical
SFXT (large and eccentric orbit), or an intermediate SFXT (small
orbits, circular or eccentric) with possible periodic outbursts \citep[see also][for an alternative model]{2007Sidoli}. 

Considering the longer duration of its flares as well as their 18.52~days
periodicity, and its lower $L_{\rm max}$/$L_{\rm min}$
ratio, IGR~J18483$-$0311 is likely an intermediate SFXT. Moreover, to figure out - in the 
framework of the model proposed 
by \citet{2008Negueruela} - whether its orbit is circular or elliptic, we can first use the 
Kepler's third law for a B0.5Ia supergiant mass and radius
$M_{\rm \ast}=33{\textrm{M}_{\rm \odot}}$ and $R_{\rm \ast}=33.8{\textrm{R}_{\rm \odot}}$ \citep{2008Searle}, 
and a neutron star mass of $1.4{\textrm{M}_{\rm \odot}}$ to derive the semi-major axis and we find 
$a\,\sim\,2.83R_{\rm \ast}$. Then, assuming that IGR~J18483$-$0311 only gets into activity 
when orbiting within the zone of high clump density, 
we find that an eccentricity $0.43 \le e \le 0.68$ is needed to explain the average three days bursting activity 
reported in \citet{2007Sguera}. We then conclude that IGR~J18483$-$0311 is an intermediate SFXT with a small and 
eccentric orbit. 
\begin{acknowledgements}
  We are grateful to J\'er\^ome Rodriguez for his useful website in
  which all \textit{INTEGRAL} sources are referenced
  (http://isdc.unige.ch/$\sim$rodrigue/html/igrsources.html). 
  F.R. acknowledges the CNRS/INSU
  for the funding of the third year of his ESO/CEA PhD studentship.
  This research has made use of NASA's Astrophysics
  Data System, of the SIMBAD and VizieR databases operated at
  CDS, Strasbourg, France, of products from the US Naval
  Observatory catalogues, of products from the Two
  Micron All Sky Survey as well as
  products from the Galactic Legacy Infrared Mid-Plane Survey
  Extraordinaire, which is a \textit{Spitzer Space Telescope} Legacy
  Science Program.
\end{acknowledgements}
\bibliographystyle{aa}
\bibliography{./mybib}{}

\begin{thebibliography}{27}
\expandafter\ifx\csname natexlab\endcsname\relax\def\natexlab#1{#1}\fi

\bibitem[{{Benjamin} {et~al.}(2003){Benjamin}, {Churchwell}, {Babler}, {Bania},
  {Clemens}, {Cohen}, {Dickey}, {Indebetouw}, {Jackson}, {Kobulnicky},
  {Lazarian}, {Marston}, {Mathis}, {Meade}, {Seager}, {Stolovy}, {Watson},
  {Whitney}, {Wolff}, \& {Wolfire}}]{2003Benjamin}
{Benjamin}, R.~A., {Churchwell}, E., {Babler}, B.~L., {et~al.} 2003, \pasp,
  115, 953

\bibitem[{{Bird} {et~al.}(2007){Bird}, {Malizia}, {Bazzano}, {Barlow},
  {Bassani}, {Hill}, {B{\'e}langer}, {Capitanio}, {Clark}, {Dean}, {Fiocchi},
  {G{\"o}tz}, {Lebrun}, {Molina}, {Produit}, {Renaud}, {Sguera}, {Stephen},
  {Terrier}, {Ubertini}, {Walter}, {Winkler}, \& {Zurita}}]{2007Bird}
{Bird}, A.~J., {Malizia}, A., {Bazzano}, A., {et~al.} 2007, \apjs, 170, 175

\bibitem[{{Bodaghee} {et~al.}(2007){Bodaghee}, {Courvoisier}, {Rodriguez},
  {Beckmann}, {Produit}, {Hannikainen}, {Kuulkers}, {Willis}, \&
  {Wendt}}]{2007Bodaghee}
{Bodaghee}, A., {Courvoisier}, T.~J.-L., {Rodriguez}, J., {et~al.} 2007, \aap,
  467, 585

\bibitem[{{Chaty} \& {Rahoui}(2007)}]{2006Chaty}
{Chaty}, S. \& {Rahoui}, F. 2007, Proceeding of the VIth INTEGRAL Workshop,
  ''The Obscured Universe'', Space Research Institute, Moscow, Russia, 2006,
  ESA's Publications Division: Special Publication SP-622

\bibitem[{{Chaty} {et~al.}(2008){Chaty}, {Rahoui}, {Foellmi}, {Tomsick},
  {Rodriguez}, \& {Walter}}]{2008Chaty}
{Chaty}, S., {Rahoui}, F., {Foellmi}, C., {et~al.} 2008, \aap, 484, 783

\bibitem[{{Chernyakova} {et~al.}(2003){Chernyakova}, {Lutovinov}, {Capitanio},
  {Lund}, \& {Gehrels}}]{2003Chernyakova}
{Chernyakova}, M., {Lutovinov}, A., {Capitanio}, F., {Lund}, N., \& {Gehrels},
  N. 2003, The Astronomer's Telegram, 157, 1

\bibitem[{{Clark} \& {Steele}(2000)}]{2000Clark}
{Clark}, J.~S. \& {Steele}, I.~A. 2000, \aaps, 141, 65

\bibitem[{{Hanson} {et~al.}(1996){Hanson}, {Conti}, \& {Rieke}}]{1996Hanson}
{Hanson}, M.~M., {Conti}, P.~S., \& {Rieke}, M.~J. 1996, \apjs, 107, 281

\bibitem[{{Hanson} {et~al.}(2005){Hanson}, {Kudritzki}, {Kenworthy}, {Puls}, \&
  {Tokunaga}}]{2005Hanson}
{Hanson}, M.~M., {Kudritzki}, R.-P., {Kenworthy}, M.~A., {Puls}, J., \&
  {Tokunaga}, A.~T. 2005, \apjs, 161, 154

\bibitem[{{in't Zand}(2005)}]{2005Zand}
{in't Zand}, J.~J.~M. 2005, \aap, 441, L1

\bibitem[{{Landolt}(1992)}]{1992Landolt}
{Landolt}, A.~U. 1992, \aj, 104, 340

\bibitem[{{Lebrun} {et~al.}(2003){Lebrun}, {Leray}, {Lavocat}, {Cr{\'e}tolle},
  {Arqu{\`e}s}, {Blondel}, {Bonnin}, {Bou{\`e}re}, {Cara}, {Chaleil}, {Daly},
  {Desages}, {Dzitko}, {Horeau}, {Laurent}, {Limousin}, {Mathy}, {Mauguen},
  {Meignier}, {Molini{\'e}}, {Poindron}, {Rouger}, {Sauvageon}, \&
  {Tourrette}}]{2003Lebrun}
{Lebrun}, F., {Leray}, J.~P., {Lavocat}, P., {et~al.} 2003, \aap, 411, L141

\bibitem[{{Levine} \& {Corbet}(2006)}]{2006Levine}
{Levine}, A.~M. \& {Corbet}, R. 2006, The Astronomer's Telegram, 940, 1

\bibitem[{{Leyder} {et~al.}(2007){Leyder}, {Walter}, {Lazos}, {Masetti}, \&
  {Produit}}]{2007Leyder}
{Leyder}, J.-C., {Walter}, R., {Lazos}, M., {Masetti}, N., \& {Produit}, N.
  2007, \aap, 465, L35

\bibitem[{{Maiolino} {et~al.}(1996){Maiolino}, {Rieke}, \&
  {Rieke}}]{1996Maiolino}
{Maiolino}, R., {Rieke}, G.~H., \& {Rieke}, M.~J. 1996, \aj, 111, 537

\bibitem[{{Masetti} {et~al.}(2008){Masetti}, {Mason}, {Morelli}, {Cellone},
  {McBride}, {Palazzi}, {Bassani}, {Bazzano}, {Bird}, {Charles}, {Dean},
  {Galaz}, {Gehrels}, {Landi}, {Malizia}, {Minniti}, {Panessa}, {Romero},
  {Stephen}, {Ubertini}, \& {Walter}}]{2008Masetti}
{Masetti}, N., {Mason}, E., {Morelli}, L., {et~al.} 2008, \aap, 482, 113

\bibitem[{{Molkov} {et~al.}(2004){Molkov}, {Cherepashchuk}, {Lutovinov},
  {Revnivtsev}, {Postnov}, \& {Sunyaev}}]{2004Molkov}
{Molkov}, S.~V., {Cherepashchuk}, A.~M., {Lutovinov}, A.~A., {et~al.} 2004,
  Astronomy Letters, 30, 534

\bibitem[{{Negueruela} {et~al.}(2006){Negueruela}, {Smith}, {Reig}, {Chaty}, \&
  {Torrej{\'o}n}}]{2006Negueruelaa}
{Negueruela}, I., {Smith}, D.~M., {Reig}, P., {Chaty}, S., \& {Torrej{\'o}n},
  J.~M. 2006, in ESA Special Publication, Vol. 604, The X-ray Universe 2005,
  ed. A.~{Wilson}, 165--170

\bibitem[{{Negueruela} {et~al.}(2008){Negueruela}, {Torrej{\'o}n}, {Reig},
  {Rib{\'o}}, \& {Smith}}]{2008Negueruela}
{Negueruela}, I., {Torrej{\'o}n}, J.~M., {Reig}, P., {Rib{\'o}}, M., \&
  {Smith}, D.~M. 2008, in American Institute of Physics Conference Series, Vol.
  1010, American Institute of Physics Conference Series, 252--256

\bibitem[{{Persson} {et~al.}(1998){Persson}, {Murphy}, {Krzeminski}, {Roth}, \&
  {Rieke}}]{1998Persson}
{Persson}, S.~E., {Murphy}, D.~C., {Krzeminski}, W., {Roth}, M., \& {Rieke},
  M.~J. 1998, \aj, 116, 2475

\bibitem[{{Rahoui} {et~al.}(2008){Rahoui}, {Chaty}, {Lagage}, \&
  {Pantin}}]{2008Rahoui}
{Rahoui}, F., {Chaty}, S., {Lagage}, P.-O., \& {Pantin}, E. 2008, \aap, 484,
  801

\bibitem[{{Searle} {et~al.}(2008){Searle}, {Prinja}, {Massa}, \&
  {Ryans}}]{2008Searle}
{Searle}, S.~C., {Prinja}, R.~K., {Massa}, D., \& {Ryans}, R. 2008, \aap, 481,
  777

\bibitem[{{Sguera} {et~al.}(2007){Sguera}, {Hill}, {Bird}, {Dean}, {Bazzano},
  {Ubertini}, {Masetti}, {Landi}, {Malizia}, {Clark}, \& {Molina}}]{2007Sguera}
{Sguera}, V., {Hill}, A.~B., {Bird}, A.~J., {et~al.} 2007, \aap, 467, 249

\bibitem[{{Sidoli} {et~al.}(2007){Sidoli}, {Romano}, {Mereghetti}, {Paizis},
  {Vercellone}, {Mangano}, \& {G{\"o}tz}}]{2007Sidoli}
{Sidoli}, L., {Romano}, P., {Mereghetti}, S., {et~al.} 2007, \aap, 476, 1307

\bibitem[{{Ubertini} {et~al.}(2003){Ubertini}, {Lebrun}, {Di Cocco}, {Bazzano},
  {Bird}, {Broenstad}, {Goldwurm}, {La Rosa}, {Labanti}, {Laurent}, {Mirabel},
  {Quadrini}, {Ramsey}, {Reglero}, {Sabau}, {Sacco}, {Staubert}, {Vigroux},
  {Weisskopf}, \& {Zdziarski}}]{2003Ubertini}
{Ubertini}, P., {Lebrun}, F., {Di Cocco}, G., {et~al.} 2003, \aap, 411, L131

\bibitem[{{Walter} \& {Zurita Heras}(2007)}]{2007Walter}
{Walter}, R. \& {Zurita Heras}, J. 2007, \aap, 476, 335

\bibitem[{{Winkler} {et~al.}(2003){Winkler}, {Courvoisier}, {Di Cocco},
  {Gehrels}, {Gim{\'e}nez}, {Grebenev}, {Hermsen}, {Mas-Hesse}, {Lebrun},
  {Lund}, {Palumbo}, {Paul}, {Roques}, {Schnopper}, {Sch{\"o}nfelder},
  {Sunyaev}, {Teegarden}, {Ubertini}, {Vedrenne}, \& {Dean}}]{2003Winkler}
{Winkler}, C., {Courvoisier}, T.~J.-L., {Di Cocco}, G., {et~al.} 2003, \aap,
  411, L1

\end{thebibliography}

\end{document}